\documentclass[aps,prl,twocolumn,superscriptaddress]{revtex4-1}
\usepackage{natbib}
\usepackage{graphicx}
\usepackage{lmodern}

\usepackage{dcolumn}
\usepackage{bm}
\usepackage{color}
\usepackage{ulem}
\usepackage[usenames,dvipsnames,svgnames,table]{xcolor}

\newcommand{\pstar}{$p^{\star}$}

\newcommand{\kzero}{$\kappa_0/T$}

\newcommand{\Kxy}{$\kappa_{xy}$}
\newcommand{\Kxx}{$\kappa_{xx}$}
\newcommand{\Kzy}{$\kappa_{zy}$}
\newcommand{\Kzz}{$\kappa_{zz}$}

\newcommand{\RH}{$R_{\rm H}$}
\newcommand{\nH}{$n_{\rm H}$}

\newcommand{\Tc}{$T_{\rm c}$}
\newcommand{\TN}{$T_{\rm N}$}
\newcommand{\Tstar}{$T^{\star}$}

\newcommand{\mstar}{$m^{\star}$}

\begin{document}

\title{Transport signatures of the pseudogap critical point in the cuprate superconductor Bi$_2$Sr$_{2-x}$La$_x$CuO$_{6+\delta}$}

\author{M. Lizaire$^{\star \star}$}
\affiliation{Institut Quantique, D\'{e}partement de physique and RQMP, Universit\'{e} de Sherbrooke, Sherbrooke, Qu\'{e}bec J1K 2R1, Canada}

\author{A. Legros$^{\star \star}$}
\affiliation{Institut Quantique, D\'{e}partement de physique and RQMP, Universit\'{e} de Sherbrooke, Sherbrooke, Qu\'{e}bec J1K 2R1, Canada}
\affiliation{SPEC, CEA, CNRS-UMR3680, Universit\'{e} Paris-Saclay, Gif-sur-Yvette Cedex 91191, France}

\author{A. Gourgout}
\affiliation{Institut Quantique, D\'{e}partement de physique and RQMP, Universit\'{e} de Sherbrooke, Sherbrooke, Qu\'{e}bec J1K 2R1, Canada}

\author{S. Benhabib}
\affiliation{Laboratoire National des Champs Magn\'{e}tiques Intenses (CNRS, EMFL, INSA, UGA, UPS), Grenoble/Toulouse, France}

\author{S. Badoux}
\affiliation{Institut Quantique, D\'{e}partement de physique and RQMP, Universit\'{e} de Sherbrooke, Sherbrooke, Qu\'{e}bec J1K 2R1, Canada}

\author{F. Lalibert\'{e}}
\affiliation{Institut Quantique, D\'{e}partement de physique and RQMP, Universit\'{e} de Sherbrooke, Sherbrooke, Qu\'{e}bec J1K 2R1, Canada}

\author{M.-E.~Boulanger}
\affiliation{Institut Quantique, D\'{e}partement de physique and RQMP, Universit\'{e} de Sherbrooke, Sherbrooke, Qu\'{e}bec J1K 2R1, Canada}

\author{A. Ataei}
\affiliation{Institut Quantique, D\'{e}partement de physique and RQMP, Universit\'{e} de Sherbrooke, Sherbrooke, Qu\'{e}bec J1K 2R1, Canada}

\author{G. Grissonnanche}
\affiliation{Institut Quantique, D\'{e}partement de physique and RQMP, Universit\'{e} de Sherbrooke, Sherbrooke, Qu\'{e}bec J1K 2R1, Canada}

\author{D. LeBoeuf}
\affiliation{Laboratoire National des Champs Magn\'{e}tiques Intenses (CNRS, EMFL, INSA, UGA, UPS), Grenoble/Toulouse, France}

\author{S. Licciardello}
\affiliation{High Field Magnet Laboratory (HFML-EMFL), Radboud University, Toernooiveld 7, Nijmegen 6525 ED, Netherlands}%

\author{S.~Wiedmann}
\affiliation{High Field Magnet Laboratory (HFML-EMFL), Radboud University, Toernooiveld 7, Nijmegen 6525 ED, Netherlands}%

\author{S.~Ono}
\affiliation{Central Research Institute of Electric Power Industry, Chiyoda-ku, Tokyo 100-8126, Japan}%

\author{H.~Raffy}
\affiliation{Laboratoire de Physique des Solides, Universit\'{e} Paris-Sud, Universit\'{e} Paris-Saclay, CNRS UMR 8502, Orsay, France}

\author{S.~Kawasaki}
\affiliation{Department of Physics, Okayama University, Okayama 700-8530, Japan}%

\author{G.-Q. Zheng}
\affiliation{Department of Physics, Okayama University, Okayama 700-8530, Japan}%
\affiliation{Institute of Physics, Chinese Academy of Sciences and Beijing National Laboratory for Condensed Matter Physics,  Beijing 100190, China}

\author{N. Doiron-Leyraud}
\affiliation{Institut Quantique, D\'{e}partement de physique and RQMP, Universit\'{e} de Sherbrooke, Sherbrooke, Qu\'{e}bec J1K 2R1, Canada}

\author{C. Proust}
\affiliation{Laboratoire National des Champs Magn\'{e}tiques Intenses (CNRS, EMFL, INSA, UGA, UPS), Grenoble/Toulouse, France}
\affiliation{Canadian Institute for Advanced Research, Toronto, Ontario M5G 1M1, Canada}

\author{L. Taillefer}
\email[]{louis.taillefer@usherbrooke.ca}
\affiliation{Institut Quantique, D\'{e}partement de physique and RQMP, Universit\'{e} de Sherbrooke, Sherbrooke, Qu\'{e}bec J1K 2R1, Canada}
\affiliation{Canadian Institute for Advanced Research, Toronto, Ontario M5G 1M1, Canada}

\date{\today}

\begin{abstract}
Five transport coefficients of the cuprate superconductor Bi$_2$Sr$_{2-x}$La$_x$CuO$_{6+\delta}$ were measured
in the normal state down to low temperature, reached by applying a magnetic field (up to 66~T) large enough to suppress superconductivity.
The electrical resistivity, Hall coefficient, thermal conductivity, Seebeck coefficient 
and thermal Hall conductivity
were measured in 
two overdoped single crystals,
with La concentration $x = 0.2$ (\Tc~$=18$~K) and $x = 0.0$ (\Tc~$=10$~K). 
The samples have dopings $p$
very close to the critical doping \pstar~where the pseudogap phase ends.
The resistivity displays a linear dependence on temperature whose slope is consistent with Planckian dissipation.
The Hall number \nH~decreases with reduced $p$, consistent with a drop in carrier density from 
$n = 1+p$ above \pstar~to $n=p$ below \pstar.
This drop in \nH~is concomitant with a sharp drop in the density of states inferred from prior NMR Knight shift measurements.
The thermal conductivity satisfies the Wiedemann-Franz law, showing that the
pseudogap phase at $T = 0$ is a metal whose fermionic excitations carry heat and charge as do conventional electrons.
The Seebeck coefficient diverges logarithmically at low temperature, a signature of quantum criticality.
The thermal Hall conductivity becomes negative at low temperature, showing that phonons are chiral in the pseudogap phase.
Given the observation of these same properties in other, very different cuprates, 
our study provides strong evidence for the universality of these five signatures of the pseudogap phase and its critical point.\\

$^{\star \star}$~{\it These authors contributed equally to this work}.

\end{abstract}

\maketitle

\section{Introduction}

Since their discovery, cuprate superconductors have captured the imagination of condensed matter physicists 
in a quest to elucidate the origin of their remarkably high critical temperature. 
A large array of experimental probes was used to scrutinize the exotic phases that emerge alongside superconductivity in these materials.
Among these, the pseudogap phase stands out for its enigmatic nature. 
There is no consensus on the nature of this phase nor its connection to superconductivity~\cite{Keimer2015}. 
It is characterized by several experimental signatures, in particular the opening of a momentum-dependent spectral gap
detected by angle-resolved photoemission (ARPES)~\cite{Hashimoto2014} and a loss of density of states detected by specific heat \cite{Lorama1998} 
and nuclear magnetic resonance (NMR) \cite{Alloul1989, Zheng2005}. 
Transport measurements in magnetic fields high enough to suppress superconductivity down to $T \simeq 0$ have unveiled the otherwise hidden properties of that phase in its ground state~\cite{proust2019}.
First, there is a drop in the carrier density at the critical doping \pstar~where the pseudogap phase ends, detected by the Hall number decreasing from $n_{\rm H} \simeq 1+p$ above \pstar~to $n_{\rm H} \simeq p$ below \pstar, in 
YBa$_2$Cu$_3$O$_y$ (YBCO)~\cite{badoux2016} 
and 
La$_{1.6-x}$Nd$_{0.4}$Sr$_x$CuO$_4$ (Nd-LSCO)~\cite{collignon2017fermi}.
Secondly, a $T$-linear resistivity 
down to $T \to 0$, the emblematic signature of quantum criticality~\cite{Taillefer2010}, 
is observed in 
Bi$_{2+x}$Sr$_{2-y}$CuO$_{6+\delta}$ (Bi2201)~\cite{Martin1990}, 
Nd-LSCO~\cite{Daou_NatPhy_2009}, 
La$_{2-x}$Sr$_x$CuO$_4$ (LSCO)~\cite{Cooper_Science_2009} 
and 
Bi$_2$Sr$_2$CaCu$_2$O$_{8+\delta}$ (Bi2212)~\cite{legros2019universal}, 
at \pstar. 
The slope of the linear regime 
is consistent with a scattering rate in the Planckian limit, namely $\hbar / \tau \simeq k_{\rm B} T$ -- 
as also found in organic, heavy-fermion and iron-based superconductors at their respective quantum critical points~\cite{legros2019universal,Bruin2013}.
Another phenomenon linked to quantum criticality is the logarithmic divergence of the specific heat at low temperature, observed in heavy-fermion metals~\cite{Lohneysen1994,Bianchi2003} and in the cuprates 
Nd-LSCO
and
La$_{1.8-x}$Eu$_{0.2}$Sr$_x$CuO$_4$ (Eu-LSCO)~\cite{michon2019thermodynamic}. 
A logarithmic divergence was also observed in the Seebeck coefficient of 
Nd-LSCO~\cite{DaouPRB2009} 
and 
Eu-LSCO~\cite{laliberte2011fermi} 
at \pstar, whereby $S/T \propto {\rm log}(1/T)$ as $T \to 0$
-- a third signature.
Fourth, the Wiedemann-Franz law -- which states that the thermal and electrical conductivities of electrons are equal in the $T=0$ limit --
was found to be valid in Nd-LSCO, both just above \pstar~(in the strange-metal state of $T$-linear resistivity) and 
below \pstar~(in the pseudogap phase)~\cite{michon2018wiedemann}. 
This tells us that the pseudogap phase is a metal whose fermionic excitations carry heat and charge as do conventional electrons. 
Finally, a large negative thermal Hall conductivity, \Kxy, has been observed in various cuprates below \pstar~\cite{Grissonnanche2019,Boulanger2020},
attributed to phonons that acquire chirality upon entering the pseudogap phase~\cite{Grissonnanche2020}.

\begin{figure}[t]
\includegraphics[width=0.98\columnwidth]{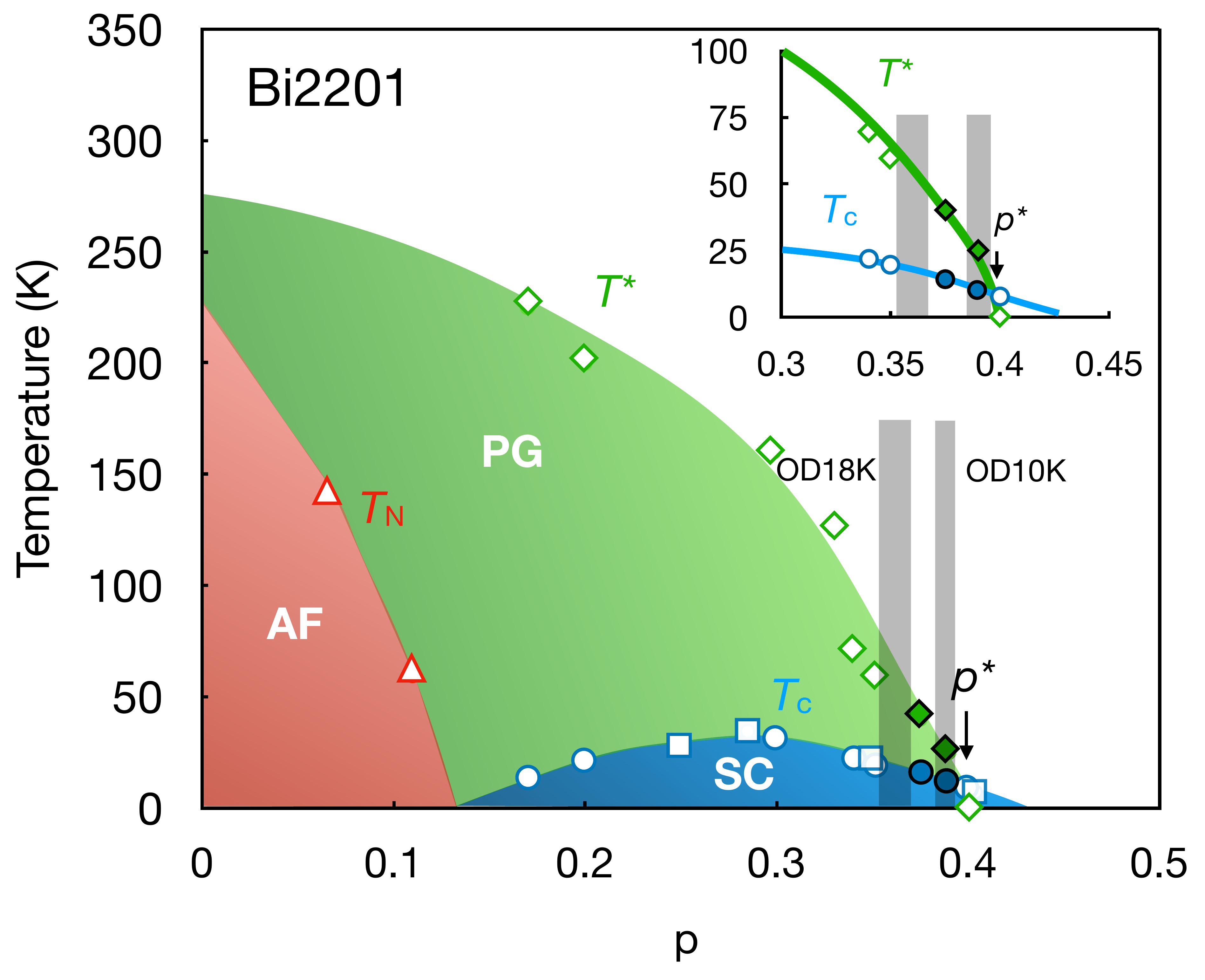}
\caption{
Temperature-doping phase diagram of the hole-doped cuprate Bi2201.
The N\'{e}el temperature \TN~(triangles), 
the superconducting temperature \Tc~(circles), and
the pseudogap temperature \Tstar~(diamonds) 
are taken from NMR measurements~\cite{kawasaki2010carrier}.
The full symbols (diamonds and circles) are from as yet unpublished data~\cite{Ito2020}.
The doping $p$ is defined from \Tc, using the experimental relation between \Tc~(squares) and the area of
the associated Fermi surface measured by ARPES~\cite{kondo2004hole}, which is proportional to $1+p$ (see text).
 The two vertical grey bands mark the dopings of our two overdoped samples,
 respectively labelled OD18K (\Tc~$=18$~K) and OD10K (\Tc~$=10$~K).
{\it Inset}:
Zoom on the region near \pstar, the critical doping where the pseudogap phase ends, {\it i.e.} where \Tstar~(green line, diamonds) goes to zero.
We see that for a sample with \Tc~$=10$~K, the pseudogap opens at \Tstar~$= 25 \pm 5$~K,
and \pstar~$= 0.40 \pm 0.01$, where \Tc~$\simeq 8$~K.
}
\label{Fig.Phase_diagram}
\end{figure}

The five transport properties outlined here have been observed all together in only one cuprate material, Nd-LSCO.
To establish that they are universal signatures of the pseudogap phase, it is necessary to confirm them in a different cuprate,
ideally all together. This is the purpose of our study, which focuses on the material Bi2201.

A number of transport studies have been reported for this material, in high fields and / or at high doping~\cite{putzke2019reduced,Ono2000,Balakirev2003,Proust2005a,Konstantinovic2002, Vedeneev2004, fruchter2007}, 
but to our knowledge, there is no prior report of 
the thermal Hall effect or the Seebeck coefficient in the normal state of overdoped Bi2201 at $T \to 0$, close to~\pstar.
This material presents multiple advantages, starting with a low maximal \Tc, 
which allows for a complete suppression of superconductivity down to $T \simeq 0$ by application of a magnetic field that can be achieved in high-field facilities~\cite{Zheng2005}.
It is a single-layer cuprate,
which facilitates the interpretation of Fermi surface properties. 
Its Fermi surface has been carefully delineated by ARPES, 
all the way to the highest dopings, beyond~\pstar~\cite{kondo2004hole,Ding2019}.
The boundary of its pseudogap phase in the temperature-doping phase diagram has been mapped out
by both ARPES~\cite{KondoNat2011} and NMR~\cite{kawasaki2010carrier}, and
the two techniques agree on the pseudogap temperature \Tstar$(p)$, and therefore on the critical doping,
located at \pstar~$= 0.40 \pm 0.01$~(Fig.~1).

Finally, its superconducting dome extends over a significantly different doping range than in other cuprates, 
namely up to 
$p \simeq 0.42$~\cite{kondo2004hole} (Fig.~1),
much higher than in YBCO, Bi2212 or LSCO, for example, where it ends at $p \simeq 0.27$~\cite{kondo2004hole}, 
thereby making Bi2201 an important candidate to test the universality of the transport signatures at \pstar. 
Furthermore, 
Bi2201 is the only hole-doped cuprate currently known to exhibit charge density wave (CDW) order {\it outside} the pseudogap phase, 
{\it i.e.} above \pstar~and well above \Tstar~\cite{Peng2018}.
Studying Bi2201 therefore allows us to separate what is due to the pseudogap from what is due to CDW order.
Bi2201 is also the only material for which NMR measurements have been carried out up to (and across)~\pstar,
in the field-induced normal state down to $T \simeq 0$~\cite{kawasaki2010carrier}, 
precisely 
where all the transport signatures of \pstar~have so far been obtained -- and where the focus of this Article will be (Fig.~1).
Studying Bi2201 therefore allows us to directly compare the transport and NMR signatures of the pseudogap phase close to \pstar.

In this Article, we report measurements of the resistivity, the Hall and Seebeck coefficients,
the thermal and thermal Hall conductivities, in the field-induced normal state of Bi2201,
at two dopings very close to \pstar. 
Our study reveals signatures of the pseudogap critical point that are very similar to those observed in Nd-LSCO: 
the carrier density drops when the doping decreases below \pstar, 
the $T$-linear dependence of $\rho(T)$ at low \textit{T} and close to \pstar~has a slope consistent with Planckian dissipation, 
the Wiedemann-Franz law holds, 
the Seebeck coefficient diverges logarithmically as $T \to 0$, 
and the thermal Hall conductivity is negative at low $T$ (as opposed to the electrical Hall conductivity).
This shows that these five 
signatures are very likely to be universal amongst hole-doped cuprates.

\begin{figure}[t]
\includegraphics[width=0.87\columnwidth]{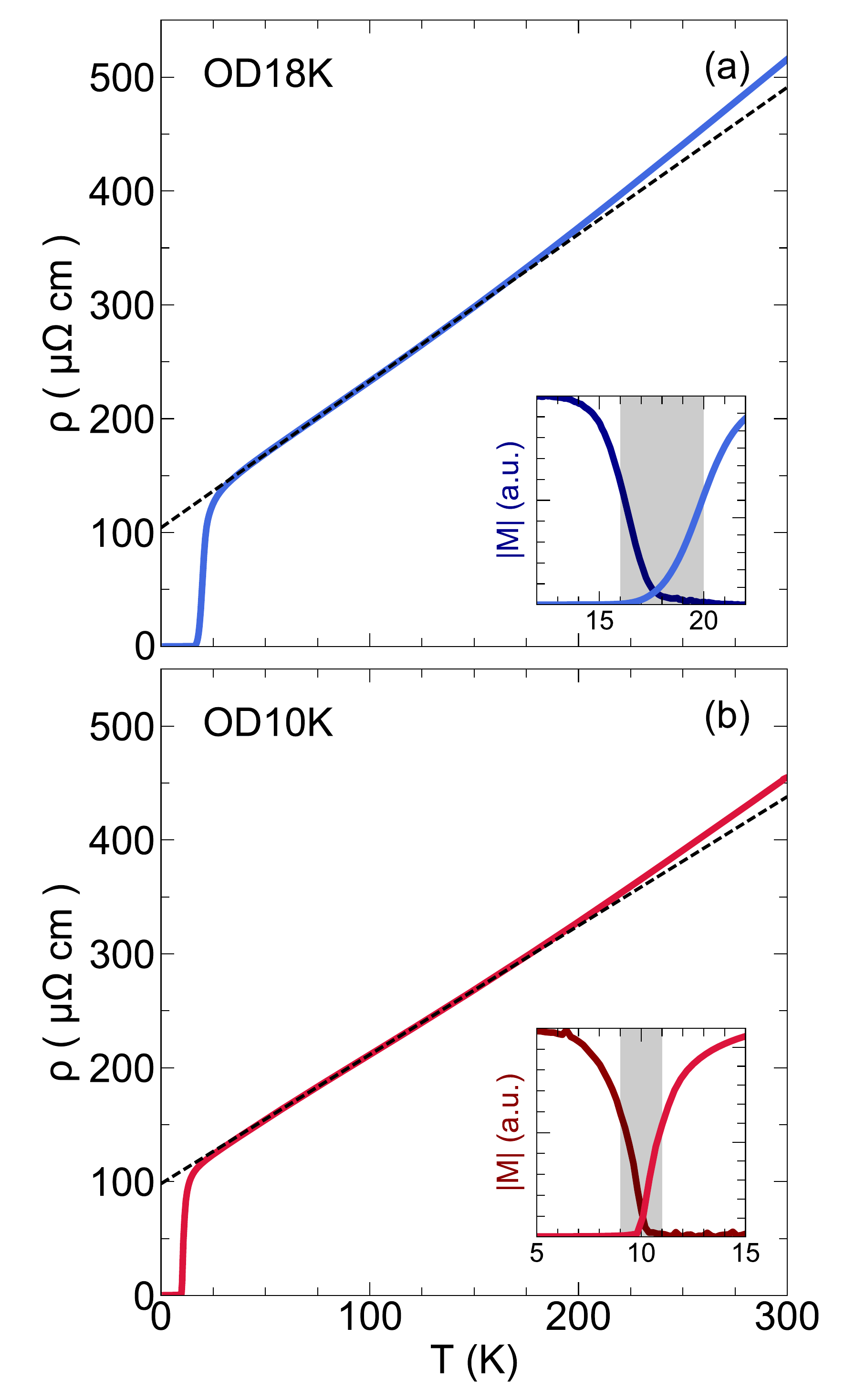}
\caption{
In-plane electrical resistivity of overdoped Bi2201 in zero magnetic field, as a function of temperature, 
for our two samples: (a) OD18K; (b) OD10K. 
The dashed line is a linear fit to the data over the interval from 60~K to 160~K.
It yields the residual resistivity $\rho_{00}$ by extrapolation to $T=0$, namely:
(a) $104 \pm 2~\mu \Omega$~cm; (b) $98 \pm 2~\mu \Omega$~cm.
{\it Inset}: 
Zoom on the magnetization (dark curve, absolute value) and the resistivity~(pale curve) near \Tc.
Together, the two curves allow us to define the bulk value of \Tc~for each sample (grey band), namely:
(a) $18 \pm 2$~K; (b) $10 \pm 1$~K.
}
\label{Fig.squid}
\end{figure}

\section{Methods}

Two single crystals, of composition Bi$_{2.05}$Sr$_{1.95}$CuO$_{6+\delta}$ and Bi$_2$Sr$_{1.8}$La$_{0.2}$CuO$_{6+\delta}$, 
were grown by the floating-zone technique \cite{ono2003evolution}. 
They are thin rectangular platelets, with a length of 2.5~mm, a width of 0.5~mm and a thickness of 0.04~mm (normal to the CuO$_2$ plane). 
Characterization by SQUID magnetization yielded sharp superconducting transitions with \Tc~$=10 \pm 1$~K 
and \Tc~$=18 \pm 2$~K (insets of Fig.~\ref{Fig.squid}). 
We label the samples OD10K and OD18K, respectively.

The resistivity $\rho$, 
Hall coefficient \RH, 
thermal conductivity $\kappa$,
Seebeck coefficient $S$,
and 
thermal Hall conductivity \Kxy were measured for both samples.
Contacts were prepared using silver epoxy, annealed at 400$^{\circ}$C for 10 minutes and quenched at room temperature. The resulting contact resistances were less than 4 Ohms at room temperature. 
The currents (electrical and thermal) were applied in the CuO$_2$ planes, 
that is along the length of the samples and the magnetic field was applied along the $c$ axis. 
Electrical transport was first performed in Sherbrooke at $H=0$ and $H=16$~T in a PPMS, 
then in pulsed magnetic fields up to $H=66$~T at the LNCMI in Toulouse,
and in a static field of $H=33$~T at the HFML in Nijmegen (for OD10K).
(In addition, we also measured the resistivity and Hall coefficient of a La-doped thin film of Bi2201 with \Tc~$\simeq 20$~K up to 66~T.)
For detailed Methods on how resistivity and Hall effect are measured in high magnetic fields, see ref.~\onlinecite{collignon2017fermi, badoux2016}.
The thermal conductivity was measured in Sherbrooke using a dilution refrigerator down to 80~mK, with an applied field up to $H=15$~T, as in ref.~\onlinecite{michon2018wiedemann} (see also ref.~\onlinecite{shakeripour2009heat} for a review on thermal conductivity measurements in superconductors).
Thermoelectricity was first measured in Sherbrooke using a cryostat with a VTI and a superconducting magnet up to $H=18$~T, 
and then at the LNCMI in Grenoble under a static field up to $H=34$~T, as in ref.~\onlinecite{collignon2021thermopower}.
The thermal Hall conductivity was measured in Sherbrooke at $H=15$~T (see ref.~\onlinecite{grissonnanche2016wiedemann, Grissonnanche2019} for experimental details; data for one of our samples, OD18K, has already been reported in ref.~\onlinecite{Grissonnanche2019}).

\paragraph{Bi$_2$Sr$_{2-x}$La$_x$CuO$_{6+\delta}$}{
As mentioned in the Introduction, the phase diagram of Bi2201 is different from many other hole-doped cuprates in the sense that the superconducting dome is located at much higher doping values, as observed in ARPES experiments. Furthermore, the question of relative doping in this cuprate is a delicate one, since there are many ways to dope the compound, {\it e.g.} La/Sr substitution \cite{kawasaki2010carrier}, Bi/Sr ratio~\cite{Vedeneev2004} and excess oxygen \cite{fruchter2007} (Bi/Pb substitution also allows to suppress the superstructure \cite{Ding2019}). These differences affect the maximum \Tc~in different ways, which makes a comparison between studies difficult, especially when several doping methods are combined. In order to connect the relative position of our La-doped samples to the pseudogap phase and \pstar, we compare the \Tc~values of our samples to the \Tc~values of the samples used in the NMR study by Kawasaki \textit{et al.}~\cite{kawasaki2010carrier}, where the same doping method was used, \textit{i.e.} La/Sr substitution. This NMR study clearly reveals a closing of the pseudogap for the La composition $x$ = 0 (\Tc~$\simeq 8$~K). The inset of Fig.~\ref{Fig.Phase_diagram} shows a zoom on the region near \pstar, including unpublished data from two additional samples (solid symbols, ref.~\onlinecite{Ito2020}): one sample has \Tc~$=10$~K and \Tstar~$= 25 \pm 5$~K; the other sample has \Tc~$=14$~K and \Tstar~$= 40 \pm 10$~K. The end of the pseudogap phase is seen to correspond to \Tc~$\simeq 8$~K. So according to this NMR-derived phase diagram, our sample OD10K is located just below \pstar. Different ways are used to define the doping $p$ in Bi2201. Here, we use the relation between \Tc~and  $p$ established from ARPES studies by Kondo {\it et al.}~\cite{kondo2004hole}, where the value of $p$ is obtained from the area of the measured Fermi surface, which is proportional to $1+p$ by the Luttinger theorem. From this relation, the end of the superconducting dome in Bi2201 is located at $p \simeq 0.42$ (Fig.~1). (A recent ARPES study~\cite{Ding2019} obtained a similar relation between \Tc~and $p$ in Pb-doped Bi2201.) Using this \Tc$-p$ relation, we obtain $p = 0.360 \pm 0.008$ for OD18K and $p = 0.390 \pm 0.005$ for OD10K (Fig.~1). It is important to note that the main conclusions of this Article do not depend on the absolute values of doping, but only on the relative position of our samples with respect to the end of the pseudogap phase (which is clear from the NMR measurements).
}

%

\begin{figure}[t]
\includegraphics[width=1\columnwidth]{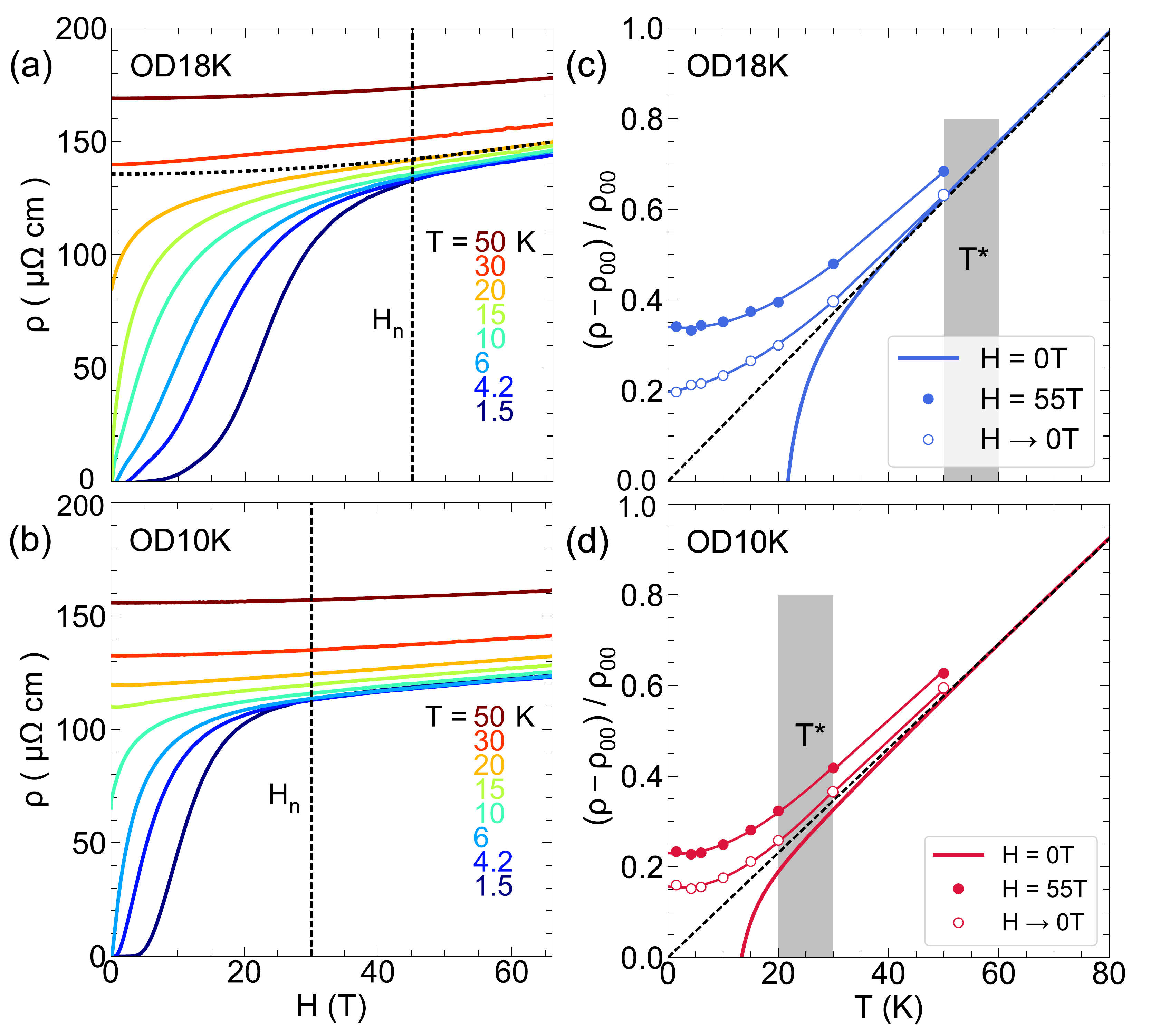}
\caption{\label{Fig.Resistivity}
{\it Left panels:}
Resistivity as a function of magnetic field 
at different temperatures, as indicated, for (a) OD18K and (b) OD10K. 
The black vertical dashed lines mark the magnetic field $H_n$ above which the normal-state resistivity is reached,
namely $H_n \simeq 45$~T for OD18K and $H_n \simeq 30$~T  for OD10K.
{\it Right panels:}
Intrinsic resistivity, $\rho(T)$ - $\rho_{00}$, normalized by the residual resistivity $\rho_{00}$, 
as a function of temperature for (c) OD18K and (d) OD10K. 
$\rho_{00}$ is obtained using a linear fit to $\rho(T)$ in zero field above \Tc~(dashed line in Fig.~\ref{Fig.squid}). 
Solid colored lines represent zero field data, full circles data taken at $H=55$~T (from isotherms in left panels)
and open circles represent data obtained in high magnetic field for which a correction has been applied to remove the magnetoresistance 
(see text).
Dashed black lines are linear fits of zero field data above \Tc~(same as dashed lines in Fig.~\ref{Fig.squid}).
The solid coloured lines that go through the data points are a guide to the eye.
In panels (c) and (d), the vertical grey band marks the pseudogap temperature \Tstar~determined by NMR Knight shift measurements 
(Fig.~1).
}
\end{figure}

\begin{figure}[t]
\centering
\includegraphics[width=0.92\columnwidth]{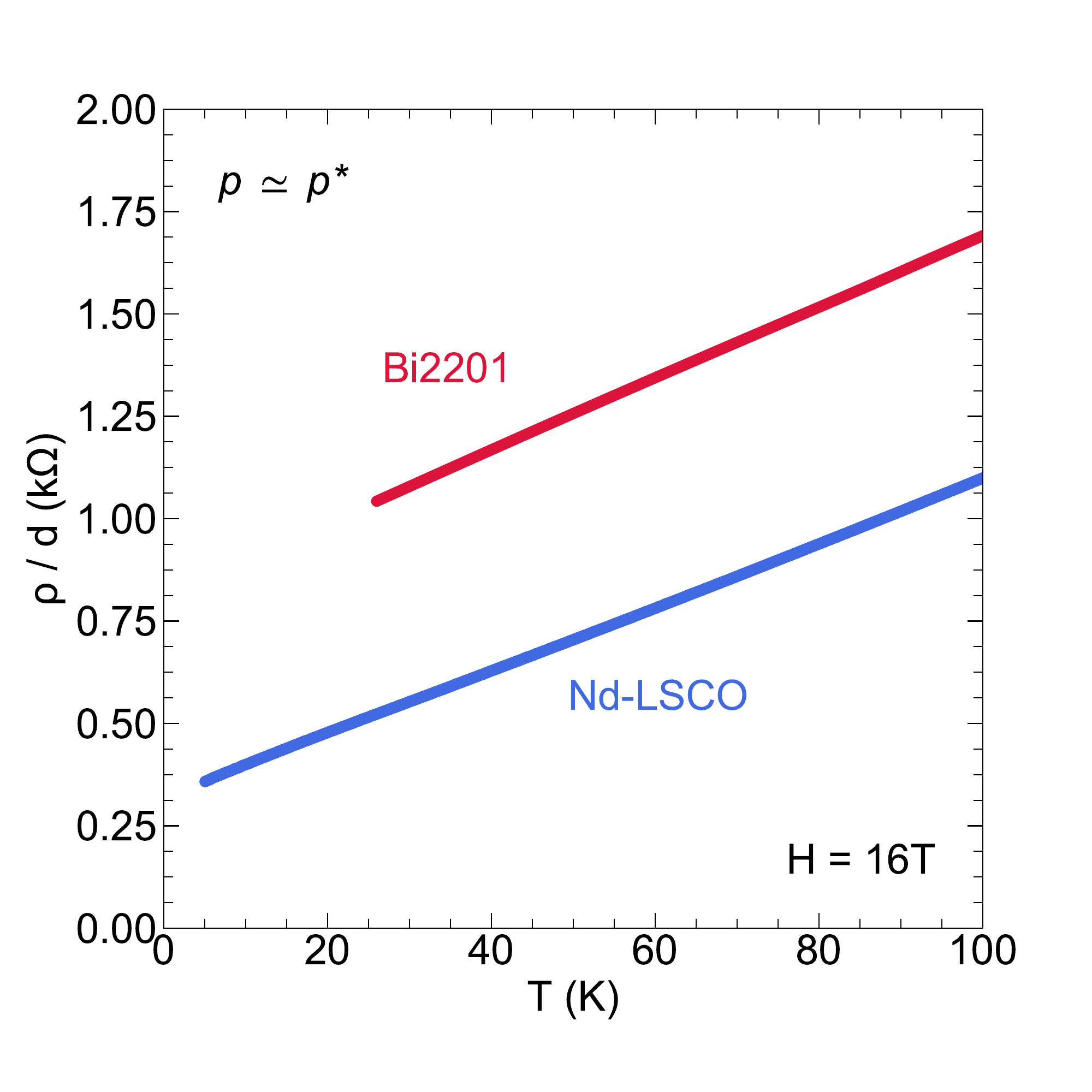}
\caption{
Normal-state resistance per CuO$_2$ plane,
defined as the resistivity $\rho$ divided by the distance $d$ between CuO$_2$ planes, as a function of temperature, 
measured in a field $H = 16$~T,
for two hole-doped cuprates close to their respective pseudogap critical points: 
Bi2201 at $p = 0.39$ (red, our OD10K sample), for which \pstar~$= 0.40$,
and Nd-LSCO at $p= 0.24$ (blue, ref.~\onlinecite{collignon2017fermi}), for which \pstar~$= 0.23$.
The two show a $T$-linear dependence,
with slopes that are very similar, namely
$A / d = 9.0 \pm 0.9~\Omega /$K
and 
$7.4 \pm 0.8~\Omega /$K~\cite{legros2019universal}, respectively. 
}
\label{Fig.Rho-vs-T_Planckian}
\end{figure}

\section{Resistivity: \textit{T}-linear dependence and Planckian dissipation}

In Fig.~\ref{Fig.squid}, the zero-field resistivity is plotted as a function of temperature up to 300~K.
The residual resistivity $\rho_{00}$ is obtained by extrapolating to $T = 0$ a linear fit to the data above \Tc~(between 60~K and 160~K). 
In Fig.~\ref{Fig.Resistivity}(c) and (d),
the zero-field resistivity is plotted as $\rho-\rho_{00}$ divided by $\rho_{00}$ (continuous curve). 
This way of plotting the data allows for an easier comparison of the inelastic resistivity in the two samples.

In Fig.~\ref{Fig.Resistivity}(a) and (b),
we show the isotherms of resistivity as a function of field up to $H=66$~T for OD18K (a) and OD10K (b), at various temperatures
from 1.5~K to 50~K.
(Isotherms obtained from our thin film of Bi2201 are very similar to those of Fig.~\ref{Fig.Resistivity}(a).)
The normal state at $T = 1.5$~K is reached when the field exceeds $H_n \simeq 45$~T for OD18K and $H_n \simeq 30$~T for OD10K. 
For both samples, we observe a small positive magnetoresistance above $H_n$. 
At $T = 50$~K, the resistivity is well described by $\rho(0) + b H^2$ over the entire field range,
where $\rho(0)$ is the resistivity at $H \to 0$.
To obtain the underlying normal state resistivity in zero field at lower temperatures,
we fit the data above $H_n$ to the form $\rho(0) + b H^2$, and extract $\rho(0)$,
as illustrated by the dotted line in Fig.~\ref{Fig.Resistivity}(a) -- the fit to the $T=20$~K isotherm of sample  OD18K.

In the right panels of Fig.~\ref{Fig.Resistivity},
we plot the $\rho(0)$ values thus obtained vs $T$ (open circles), and
compare them with the data taken at $H = 55$~T (solid circles).
The difference between the two is due to the magnetoresistance, seen to grow with decreasing $T$.

In Fig.~\ref{Fig.Resistivity}(d),
we see that in the absence of superconductivity, the resistivity free of magnetoresistance continues to be linear in $T$
down to about 20~K.
This is also seen in Fig.~\ref{Fig.Rho-vs-T_Planckian}, where $\rho(T)$ at $H = 16$~T 
is linear down to $T \simeq 25$~K.
However, the linearity does not persist as $T \to 0$, 
whereas it does in Nd-LSCO at $p = 0.24$ (Fig.~\ref{Fig.Rho-vs-T_Planckian}).
The difference is that $p >$~\pstar~in Nd-LSCO whereas $p <$~\pstar~for Bi2201.
In Nd-LSCO at $p <$~\pstar, an upturn at low $T$ develops \cite{Daou_NatPhy_2009,collignon2017fermi},
and the temperature below which the resistivity deviates upward 
from its high-temperature linear behaviour corresponds to the temperature \Tstar~measured by ARPES for the opening of the anti-nodal pseudogap
\cite{Matt2015}.
In Fig.~\ref{Fig.Resistivity}(d),
we observe the same correspondance for Bi2201.
Indeed, for 
our sample with \Tc~$=10$~K, the upward deviation in the resistivity begins
at the temperature \Tstar~measured by NMR for the onset of the pseudogap phase (grey band) \cite{Ito2020} (Fig.~1).
As we lower the doping to OD18K, 
the upward deviation gets more pronounced (Fig.~\ref{Fig.Resistivity}(c)).
Such upward deviations from linearity in $\rho(T)$ are observed not only in
Nd-LSCO~\cite{collignon2017fermi}, as mentioned above, but also in 
LSCO~\cite{boebinger1996insulator,LaliberteMIC,AndoPRL2004}.
In both cases, they are linked to the onset of the pseudogap phase, in the sense that
the deviations are seen to start at the pseudogap temperature \Tstar~measured 
by ARPES~\cite{Matt2015,collignon2017fermi,Yoshida2009,Cyr-Choiniere2018}.
They are attributed to a drop in carrier density that starts below \Tstar.
Note that in LSCO the presence of short-range magnetism below \pstar~\cite{Frachet2020},
at low temperature ($T \ll$~\Tstar), also plays a role in causing $\rho(T)$ to increase as $T \to 0$~\cite{Bourgeois-Hope2019}.

In Bi2201, the signature of \Tstar~in the resistivity depends on the magnitude of \Tstar. 
When \Tstar~is large, at low doping, the resistivity shows a downward deviation below \Tstar~from its $T$-linear dependence at high $T$~\cite{KondoNat2011},
as in YBCO~\cite{AndoPRL2004,Cyr-Choiniere2018}.
When \Tstar~is small, at high doping, an upward deviation is seen below \Tstar,
as in Nd-LSCO and LSCO~\cite{collignon2017fermi,Cyr-Choiniere2018}.
The former effect can be attributed to a loss of inelastic scattering.
The latter effect has been attributed to a loss of carrier density,
with the possible added role of scattering from short-range magnetism.
In the case of Bi2201 (with Pb substitution), the development of Cu-spin correlations has been reported below 2K across the superconducting regime~\cite{Adachi2011} as well as the presence of ferromagnetic fluctuations in the heavily overdoped compounds~\cite{Kurashima2018}. In addition to the loss of carrier density, these magnetic correlations could contribute to the upward deviations observed at low $T$ in Figs.~\ref{Fig.Resistivity}(c) and~\ref{Fig.Resistivity}(d) \cite{Bourgeois-Hope2019}.

%
%
%
%
%

To summarize, the loss of inelastic scattering below \Tstar~is the dominant effect in underdoped YBCO and Bi2201, 
whereas in LSCO and Nd-LSCO, the loss of carrier density dominates 
(see ref.~\onlinecite{Cyr-Choiniere2018}).
As \Tstar~decreases, with increasing $p$, the strength of inelastic scattering at \Tstar~weakens and the former effect gets smaller~\cite{KondoNat2011},
becoming insignificant just below \pstar, as seen in our Bi2201 data
(and in prior Bi2201 data at high doping~\cite{Ono2000}).

It is instructive to examine the slope of the $T$-linear resistivity we observe in our Bi2201 samples. In Fig.~\ref{Fig.Rho-vs-T_Planckian}, we compare the resistivity per plane in OD10K with that of another cuprate near \pstar: Nd-LSCO at $p$~=~0.24~\cite{collignon2017fermi} (estimating the resistivity per CuO$_2$ plane allows us to compare cuprates with different crystal structures).
The two systems have very similar slopes, namely $A$/$d$~=~9.0~$\pm$~0.9~$\Omega$/K in Bi2201 and $A / d = 7.4 \pm 0.8~\Omega /$K in Nd-LSCO~\cite{legros2019universal}, where $A$ is the slope of $\rho(T)$ and $d$ is the separation between CuO$_2$ planes. A very similar slope is also found in the bilayer cuprate Bi2212 at $p = 0.23 \simeq$~\pstar, namely $A / d = 8.0 \pm 0.9~\Omega /$K~\cite{legros2019universal}. Assuming that the scattering rate has the Planckian form and magnitude, namely $\hbar / \tau = k_{\rm B} T$, and using a simple Drude form ($\rho =$~\mstar $/ n e^2 \tau$), we can estimate the theoretical value for $A$ (and therefore for $A/d$) in the hypothesis that the Planckian limit is reached: $A/d$~=~($m^*k_B$)/($ne^2\hbar d$) (see ref.~\onlinecite{legros2019universal} for a detailed analysis). In ref.~\onlinecite{legros2019universal}, for Bi2201 at \pstar, the Planckian value for the resistivity's slope was estimated to be $A/d$~$\simeq$~8$\pm$2~$\Omega$/K (to compare with the experimental value 9.0$\pm$0.9~$\Omega$/K), using specific heat data to estimate an average effective mass \mstar.
(Note that in the paper by Legros \textit{et al}. the value of the specific heat had to be extrapolated to higher magnetic fields than measured in ref.~\onlinecite{ikuta2003low} in order to extract the normal-state electronic specific heat at $H_{c2}$, yielding $\gamma$~=~10$\pm$2~mJ/K$^{-2}$mol, but recent specific heat data obtained in high magnetic fields on our sample OD10K show a good agreement with the extrapolated value: $\gamma$~=~8$\pm$1~mJ/K$^{-2}$mol at 3K~\cite{girod2021normal}. With this last value, we get a theoretical slope $A/d$~=~6.3$\pm$1.1$\Omega$/K.) Although this is a rather rough estimate, it does show that sample OD10K exhibits a behaviour consistent with Planckian dissipation (above \Tstar $\simeq 25$~K). Note that the whole notion of Planckian dissipation has recently been made much more compelling by the direct extraction of the scattering time $\tau$ from angle-dependent magneto-resistance (ADMR) measurements on Nd-LSCO at $p$~=~0.24, whereby  $1 / \tau$ is found to have a perfectly linear $T$ dependence, with a slope close to $k_{\rm B} / \hbar$~\cite{GrissonnancheADMR2020}.
There is little doubt that the same 
scattering mechanism is at play in the three cuprates mentioned here,
namely Nd-LSCO, Bi2201 and Bi2212.
There may even be a universal character to the $T$-linear resistivity seen in various metals as $T \to 0$~\cite{Bruin2013,legros2019universal}.

Of course, there have been previous reports of a $T$-linear resistivity in Bi2201,
{\it e.g.} in a crystal with \Tc~=~7~K~\cite{Martin1990} (whose doping was controlled with the ratio Bi/Sr)
and in a thin film (doped with oxygen) with \Tc~$= 8$~K~\cite{fruchter2007}. 
In both cases, the linearity was 
from room temperature down to \Tc~at $H=0$,
with a slope comparable to ours.
%

\section{Hall coefficient: drop in carrier density}

\begin{figure}[t]
\includegraphics[width=1\columnwidth]{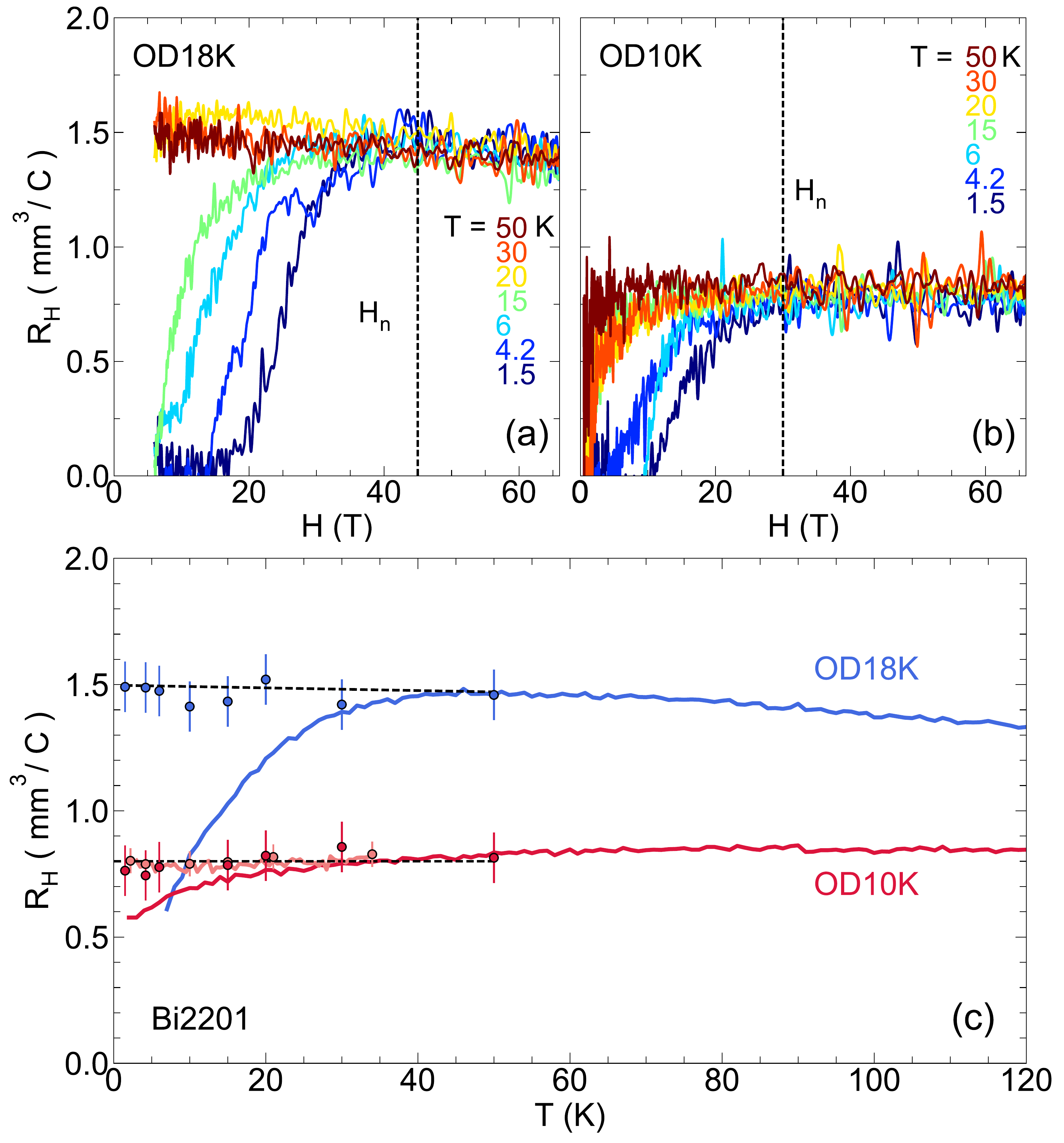}
\caption{\label{Fig.Hall}
{\it Top panels:}
Hall coefficient of Bi2201 as a function of magnetic field at different temperatures as indicated,
for (a) OD18K and (b) OD10K.
The black vertical dashed lines mark the magnetic field $H_n$ above which the normal-state Hall coefficient is reached,
namely $H_n \simeq 45$~T for OD18K and $H_n \simeq 30$~T  for OD10K. 
{\it Bottom panel:}
Hall coefficient as a function of temperature (c), 
for OD18K (blue) and OD10K (red), 
at $H=16$~T (solid lines), $H=33$~T (pink, OD10K) 
and $H=55$~T (dots with error bars, from top panels). 
The black dashed lines are a guide to the eye through the 55~T data points.
They yield the following values at $T \to 0$:
\RH~$=1.5 \pm 0.2$ mm$^3$/C for OD18K
and
$0.8 \pm 0.1$~mm$^3$/C for OD10K.
}
\end{figure}

The Hall coefficient \RH~of samples OD18K and OD10K is presented in Fig.~\ref{Fig.Hall} as a function of field up to $H$~=~66~T, 
at various temperatures. 
\RH~is almost field independent above 
$H_n$~$\sim$~45~T
in OD18K and 
$H_n$~$\sim$~30~T
in OD10K, 
in agreement with the $\rho$ vs $H$ data (Fig.~\ref{Fig.Resistivity}). 
In Fig.~\ref{Fig.Hall}(c), 
we show \RH~as a function of temperature at $H=16$~T (line),
$H=33$~T (pale line 
and circles, for OD10K)
and $H=55$~T (circles). 
We see that \RH~at $T \to 0$ jumps 
by a factor $\sim 2$ with decreasing $p$,
from $0.8 \pm 0.1$~mm$^3$/C in OD10K to $1.5 \pm 0.2$~mm$^3$/C in OD18K.
(The Hall resistance in our thin film with $T_c$~$\simeq$~20K also reaches $\sim$1.5-1.6~mm$^3$/C at low temperature.)

\begin{figure}[t]
\centering
\includegraphics[width=0.95\columnwidth]{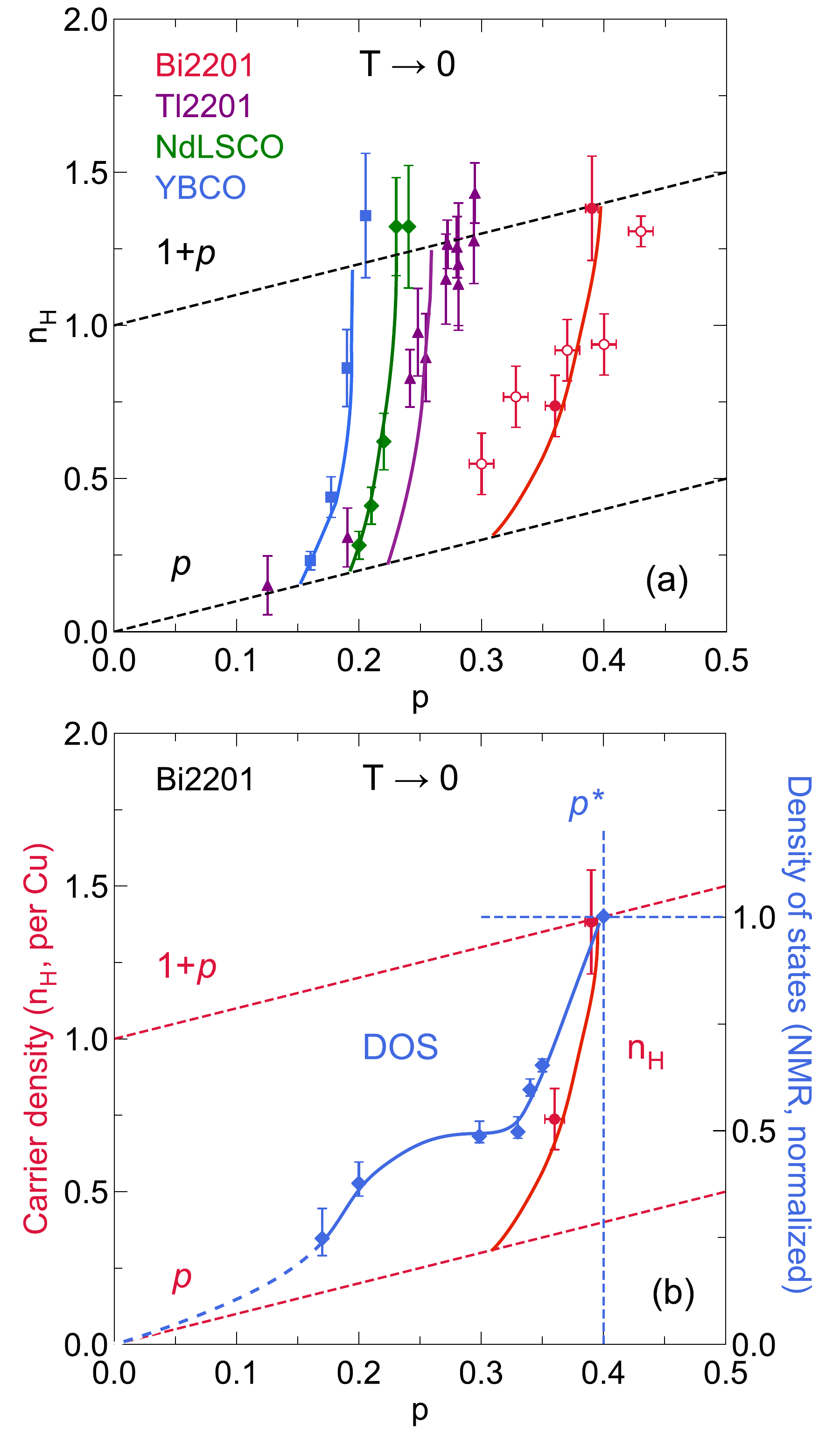}
\caption{
(a)
Hall number \nH~$= V / (e$\RH),
with \textit{V} the volume per Cu, as a function of doping for four different hole-doped cuprates: 
Bi2201 (solid red circles, this work; open red circles, ref.~\onlinecite{putzke2019reduced}),
Tl2201 (purple triangles, ref.~\onlinecite{putzke2019reduced}), 
Nd-LSCO (green diamonds, ref.~\onlinecite{collignon2017fermi}), 
and YBCO (blue squares, ref.~\onlinecite{badoux2016}).
Dashed black lines correspond to \nH~$=p$ and \nH~$=1+p$, 
and colored solid lines are a guide to the eye for each compound. 
In all four cuprates, 
we observe a drop of \nH~as doping decreases.
In Bi2201, Nd-LSCO and YBCO, the start of this drop in \nH~coincides with the onset of the pseudogap phase at \pstar,
where \pstar~is known independently:
\pstar~$= 0.4$ for Bi2201 \cite{kawasaki2010carrier,kondo2004hole}, 
\pstar~$= 0.23$ for Nd-LSCO \cite{collignon2017fermi,Matt2015}
and \pstar~$= 0.19$ for YBCO \cite{tallon2001doping}.
In Tl2201, there is currently no other measurement of \pstar.
(b)
Comparison of \nH~vs $p$ (red, left axis) 
with the density of states measured by the normal-state NMR Knight shift at $T \to 0$ (blue, right axis,~\cite{kawasaki2010carrier}),
normalized to unity at $p = 0.4$.
} 
\label{Fig.nH-vs-p}
\end{figure}

In Fig.~\ref{Fig.nH-vs-p}(a),
we plot the Hall number, \nH~$=V / (e$\RH), as a function of doping.
For a single-band metal with isotropic Fermi surface and scattering rate,
\nH~$= n$, the carrier density, in the low-temperature limit.
In Bi2201, a single-band metal, $n = 1 + p$ is the value expected from the band structure,
given by the area of the hole-like Fermi surface measured by ARPES (up to the van Hove point located at $p_{\rm vH} \simeq 0.41$)~\cite{kondo2004hole}.
In OD10K, we see that \nH~$\simeq 1 + p$, as also found in Nd-LSCO at $p = 0.24$~\cite{Daou_NatPhy_2009,collignon2017fermi} 
and Tl2201~\cite{putzke2019reduced} at $p = 0.3$~(Fig.~\ref{Fig.nH-vs-p}(a)).
Reducing the doping to OD18K, 
we find that \nH~undergoes a rapid and pronounced drop. 
%

In Fig.~\ref{Fig.nH-vs-p}(a),
we compare our data on Pb-free samples of Bi2201 (full red circles) 
to the data of Putzke \textit{et al.} on (La,Pb)-Bi2201 samples \cite{putzke2019reduced} (open red circles), 
using the same \Tc-$p$ relation to define the doping of their samples. 
We see that their data are reasonably consistent with our own, 
in particular they also show a drop in \nH~with decreasing $p$,
roughly from $1+p$ to $p$.
The slower decrease in \nH~seen in this plot of their data compared to our data could be due to the inadequacy
of applying the same $T_c-p$ relation to both studies.
(Note that Putzke {\it et al.} use a different approach to estimate the doping $p$, 
unrelated to the volume of the Fermi surface, and a different approach to define the location of \pstar, 
relying on resistivity data rather than NMR data.)

 As seen from Fig.~\ref{Fig.nH-vs-p}(a),
 the drop in \nH~we report for Bi2201 is very similar to that
 previously seen in YBCO \cite{badoux2016} and Nd-LSCO \cite{collignon2017fermi},
at their own critical doping, \pstar~$= 0.19$ and 0.23, respectively.
The observation of the same drop in \nH~below \pstar~in three different cuprate materials shows
that this signature is a robust characteristic of the critical point where the pseudogap phase ends.
We attribute it to a drop in carrier density, from $n = 1+p$ to $n = p$.
Note that this drop in carrier density has a more pronounced effect on $\rho(T)$ in Nd-LSCO than it does in Bi2201,
a difference that remains to be understood.

A drop in \nH~with decreasing $p$ has also been reported for the single-layer cuprate Tl2201~\cite{putzke2019reduced},
a material for which the location of \pstar~is still unknown.
We reproduce those data in Fig.~\ref{Fig.nH-vs-p}(a) (purple triangles).
Again, we see that \nH~drops from $1+p$ to $p$.
There is little doubt that the same fundamental mechanism is at play in all four cuprates, clearly associated with the onset of the pseudogap phase in the first three materials, suggesting that \pstar~$\simeq 0.26$ in Tl2201.

We point out that the Fermi surface of Bi2201 undergoes a Lifshitz transition at the van Hove point in its band structure,
located at the doping $p_{\rm vH} \simeq 0.41$ (\Tc~$\simeq 0$),
whereby it goes from being hole-like below that doping to being electron-like above~\cite{kondo2004hole}.
The same is true for Nd-LSCO at $p_{\rm vH} \simeq 0.23$~\cite{Matt2015} and LSCO at $p_{\rm vH} \simeq 0.18$~\cite{Yoshida2009}.
Naively, one might expect the Hall coefficient to be negative at $p > p_{\rm vH}$, but this is not the case because
the anti-nodal regions where the Fermi surface undergoes its change to an electron-like curvature contribute very little to the conductivity,
since the Fermi velocity is vanishingly small there, and the scattering is large.
A recent ADMR study of Nd-LSCO at $p = 0.24$ shows that to be the case, and explains why \RH~$>0$ even though 
$p > p_{\rm vH}$, and why \nH~$\simeq 1+p$~\cite{GrissonnancheADMR2020}. 
Similarly, in our Bi2201 sample OD10K, even if $p = 0.39$ is only slightly below $p_{\rm vH}$, 
we again get \nH~$\simeq 1+p$.
%




As shown in Fig.~\ref{Fig.nH-vs-p}(b),
the drop in \nH~below \pstar~in Bi2201 is accompanied by another clear signature observed in Bi2201, 
obtained from measurements of the NMR Knight shift~\cite{kawasaki2010carrier}: 
an abrupt drop in the $T=0$ spin susceptibility (in the absence of superconductivity) upon crossing 
below \pstar, interpreted as a drop in the normal-state density of states.
The combination of these two signatures of \pstar,
seen in the same material for the first time,
sheds new light on the nature of the 
pseudogap phase. 
The NMR data suggest that the density of states 
drops by a factor 2 upon entering the pseudogap phase,
in going from $p >$~\pstar~to $p <$~\pstar~(Fig.~\ref{Fig.nH-vs-p}(b)).
(It decreases further, towards zero, as $p \to 0$.)
A drop in the density of states by a factor 2 and a drop in the carrier density from $n = 1+p$ to $n = p$
are two properties of a transition into a phase of long-range antiferromagnetic order, with wavevector $Q = (\pi, \pi)$~\cite{Verret2017}.
Although commensurate long-range antiferromagnetic order is seen only at dopings much lower than \pstar, 
whether in Bi2201 (Fig.~1) or in any other cuprate, these experimental signatures nonetheless point to some similarity
between the pseudogap phase and the antiferromagnetic phase, perhaps in terms of short-range antiferromagnetic correlations.
A recent ADMR study of Nd-LSCO shows that the Fermi surface is transformed upon entering the pseudogap phase~\cite{Fang2020},
in a way that is consistant with small nodal hole pockets 
akin to those expected theoretically for a Fermi surface reconstruction controlled by the $Q = (\pi, \pi)$ antiferromagnetic wavevector~\cite{Storey2016,Verret2017} and recently detected experimentally (via ARPES and quantum oscillations)
in the antiferromagnetic phase of a five-layer cuprate at low doping~\cite{Kunisada2020}.

It has been suggested theoretically that charge order (or charge correlations)~\cite{Grilli-CDW-Hall}, 
or related nematic distortions~\cite{Kivelson-nematic-Hall},
may play a role in, or be responsible for, the drop in \nH~below~\pstar.
Our data in Bi2201 allow us to rule out such scenarios, given the charge order observed in overdoped Bi2201 via X-ray diffraction~\cite{Peng2018}.
Indeed, charge order is present both immediately below \pstar, in samples with \Tc~$=11$~K and 17~K, 
and immediately above \pstar, in a sample with \Tc~$=0$~K.
At $p <$~\pstar, charge order persists well above \Tstar, up to at least 250~K.
We see that charge order in Bi2201 is a completely distinct phenomenon from the pseudogap phase.
Because the same charge correlations are observed
at two dopings (\Tc~$=11$~K and 17~K)~\cite{Peng2018} that very nearly correspond to the dopings in our samples (\Tc~$=10$~K and 18~K), 
these charge correlations cannot play a key role in the pronounced change we see in the Hall coefficient between OD10K and OD18K.

\begin{figure}[t]
\includegraphics[width=1\columnwidth]{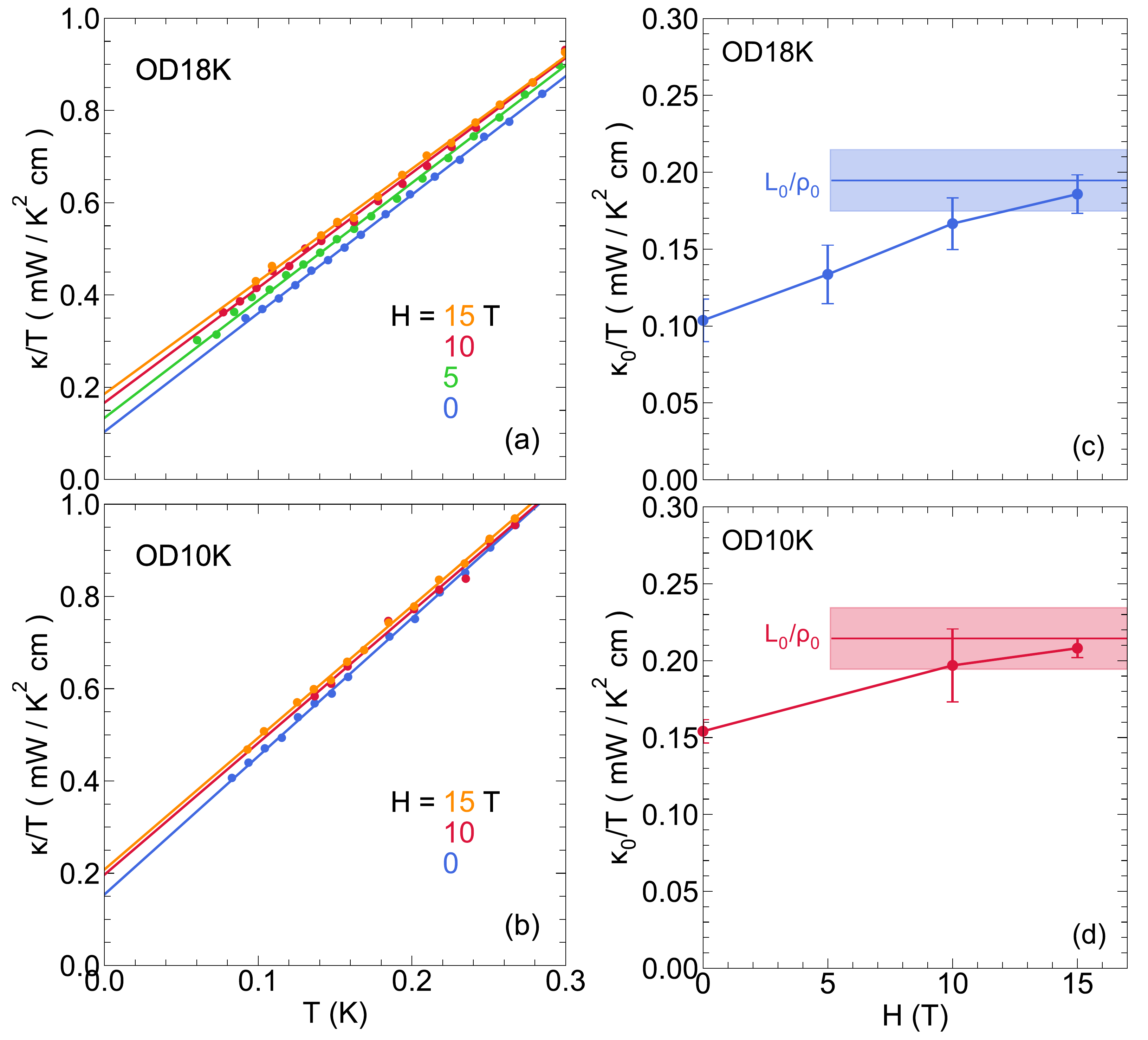}
\caption{\label{Fig.Kappa} 
{\it Left panels:} 
Thermal conductivity as a function of temperature, plotted as $\kappa/T$ vs $T$ for OD18K (a) and OD10K (b), 
at different magnetic fields, as indicated.
Solid lines are linear fits to the data, extrapolating to \kzero~at $T=0$, the electronic term.
{\it Right panels:} 
\kzero~as a function of magnetic field for OD18K (c) and OD10K (d). 
The horizontal lines correspond to the values of \kzero~expected from the Wiedemann-Franz law, 
\textit{i.e.} $\kappa_0$/\textit{T} = $L_0$/$\rho_0$, 
with $\rho_0$ the resistivity in the $T \rightarrow 0$ limit (taken at $H = 15$~T) 
and $L_0$ the Lorenz constant (error bars are shown as shaded bands).
}
\end{figure}

\section{Thermal conductivity: Wiedemann-Franz law}

Fig.~\ref{Fig.Kappa} presents thermal conductivity measurements in OD18K and OD10K at  $H=0,5,10$ and 15~T. 
In the left panels,
we plot $\kappa/T$ as a function of temperature so that the residual linear term $\kappa_0/T$ corresponds to the electronic contribution. 
The linear increase of $\kappa/T$ vs $T$ -- also observed in overdoped Nd-LSCO~\cite{michon2018wiedemann} and Tl2201~\cite{Hawthorn2007} -- 
is due to phonons that are predominantly scattered by electrons.
Data at $H=10$~T and $H=15$~T are practically superimposed, which indicates that the normal state is very nearly reached in the bulk. 
(Measurements of the specific heat on the same two samples show that the critical field for OD10K is indeed 15~T, while it is $\sim 25$~T for OD18K
\cite{girod2021normal}.)
This is reminiscent of what was found in Nd-LSCO, in which the normal state in thermal conductivity measurements was also reached at a smaller field than in electric transport measurements \cite{michon2018wiedemann}. 
This can be understood from the fact that thermal measurements are sensitive to the bulk and cannot be short-circuited by a small superconducting portion of the sample (due to an inhomogeneous doping), as can happen in electrical measurements.

Comparing the residual linear term in the superconducting state ($H=0$~T) vs the normal state ($H=15$~T), for OD10K, 
we get a ratio of $\kappa_{\rm S}/\kappa_{\rm N} \simeq 0.7$. 
The same ratio is observed in Nd-LSCO at $p=0.24$~\cite{michon2018wiedemann}, where \Tc~$\simeq 11$~K~\cite{michon2019thermodynamic}. 
Given the similar \Tc~values in the two samples, and thus a similar superconducting gap, 
this indicates a similar level of pair breaking by impurities, both compounds being in the dirty limit.  
(Note, however, that the residual resistivity per CuO$_2$ plane is twice as large in Bi2201, 
namely $\rho_{00} / d = 98~\mu \Omega$~cm~$/~12.3~\AA = 800~\Omega$, 
compared to $\rho_{00} / d = 21~\mu \Omega$~cm~$/~6.64~\AA = 320~\Omega$ in Nd-LSCO.
This suggests that the impurities/defects responsible for elastic scattering in the normal state are less effective
at breaking Cooper pairs in Bi2201 than they are in Nd-LSCO.)

In the normal state, we can test the Wiedemann-Franz law -- a hallmark of metallic behavior -- for each sample. 
This relation between charge conduction and heat conduction in the $T = 0$ limit is given by 
$\kappa_0 / T = L_0 / \rho_0$, 
where $\rho_0$ is the resistivity in the $T = 0$ limit (see below) and $L_0 = (\pi^2/3)(k_{\rm B} / e)^2$ is the Lorenz number. 
It was found to be satisfied in Nd-LSCO both below and just above \pstar~\cite{michon2018wiedemann},
and in strongly overdoped Tl2201~\cite{Proust2002}.
In the right panels of Fig.~\ref{Fig.Kappa},
we plot the electronic residual linear term \kzero~as a function of magnetic field, 
along with the value expected from the Wiedemann-Franz law, namely $L_0$/$\rho_0$, displayed as a horizontal line (with error bar). 
We estimate $\rho_0$ at $H = 15$~T,
by using the $\rho$($H$)~$\propto$~$H^2$ fits to the high field isotherms in Fig.~\ref{Fig.Resistivity} (used to correct for the magneto-resistance, as presented with the dotted black curve in OD18K), and then take a cut of these normal state isotherms at $H = 15$~T.
We then plot the temperature dependence of $\rho$($H \to 15$~T) and extrapolate to $T = 0$.
The uncertainty on the geometric factors associated with the use of
different contacts for the electrical and thermal measurements (since contacts were remade), 
leads to
an error bar of about $\pm 10$\% on $L_0$/$\rho_0$.
As we increase the field to reach the normal state, the electronic residual linear term \kzero~tends towards its electrical counterpart in both samples, 
thereby showing that, within error bars, the Wiedemann-Franz law is satisfied in Bi2201, just below \pstar.
%

\begin{figure}[t]
\includegraphics[width=1\columnwidth]{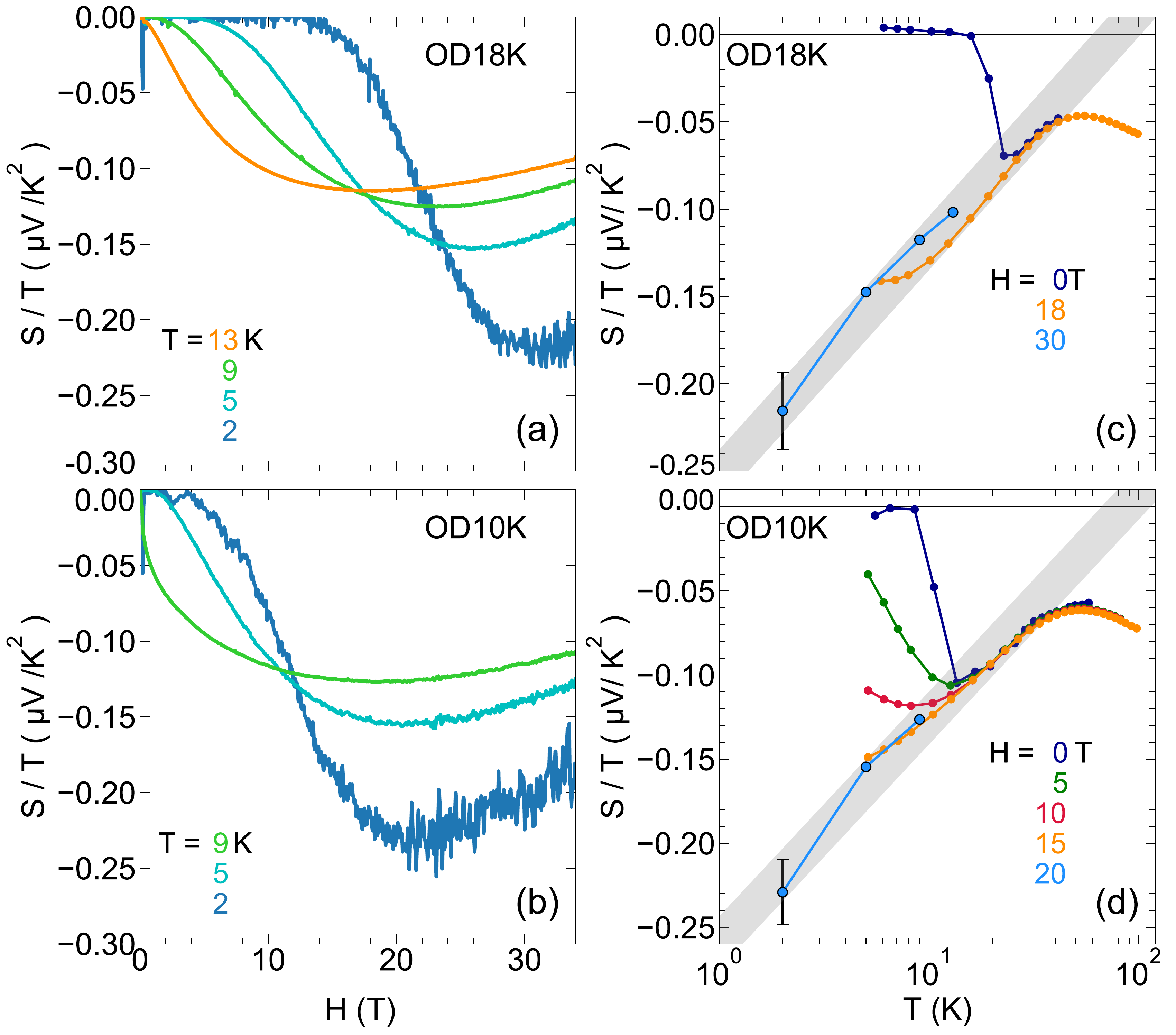}
\caption{\label{Fig.Seebeck}
{\it Left panels:} 
Seebeck coefficient as a function of magnetic field , plotted as $S/T$ vs $H$ for OD18K (a) and OD10K (b), 
at different temperatures, as indicated.
{\it Right panels:} 
Seebeck coefficient as a function of temperature, plotted as $S/T$ vs $T$ for OD18K (a) and OD10K (b), 
at different magnetic fields, as indicated.
The grey band highlights the logarithmic temperature dependence of $S/T$ observed in the normal state as $T \to 0$, 
once superconductivity is suppressed by the magnetic field.
}
\end{figure}

\section{Seebeck coefficient: logarithmic divergence}

To our knowledge, the Seebeck coefficient $S$ of Bi2201 has only been measured in zero magnetic field~\cite{Konstantinovic2002, kondo2005,Sakamoto2016},
and therefore its normal-state behaviour in the $T \to 0$ limit has hitherto been unknown.
At temperatures above \Tc, all studies find that $S$ is negative at high doping and positive at low doping.
At optimal doping (maximal \Tc), $S(T)$ goes from positive at low $T$ to negative at high $T$~\cite{kondo2005,Sakamoto2016}.
Although one might have expected $S$ to have the same sign as \RH, which is always positive in Bi2201,
calculations for a typical cuprate band structure show that a negative $S$ is consistent with a positive \RH~in the approximation
of an isotropic scattering time (but not in the approximation of an isotropic mean free path),
both below and above the van Hove point $p_{\rm vH}$~\cite{Verret2017}.
So the signs of $S$ and \RH~in Bi2201 at high doping are both consistent with the band structure.

In Fig.~\ref{Fig.Seebeck},
we report our measurements of the Seebeck coefficient in Bi2201, for our two samples.
Left panels show isotherms of $S/T$ vs $H$ up to $H = 34$~T. 
At the lowest temperature ($T=2$~K),
we see that the normal state is reached in high fields for both OD18K and OD10K, by $H \simeq 30$~T and $H \simeq 20$~T, respectively.
Right panels show the temperature dependence of $S$, plotted as $S/T$ vs log$(T)$,
combining data at low $T$ (blue dots) 
-- taken from the isotherms at $H = 30$~T (OD18K) and at $H = 20$~T (OD10K) --
with data at higher $T$ (orange dots) 
-- taken at $H = 18$~T (OD18K) and at $H = 15$~T (OD10K).
%

\begin{figure}[t]
\centering
\includegraphics[width=0.9\columnwidth]{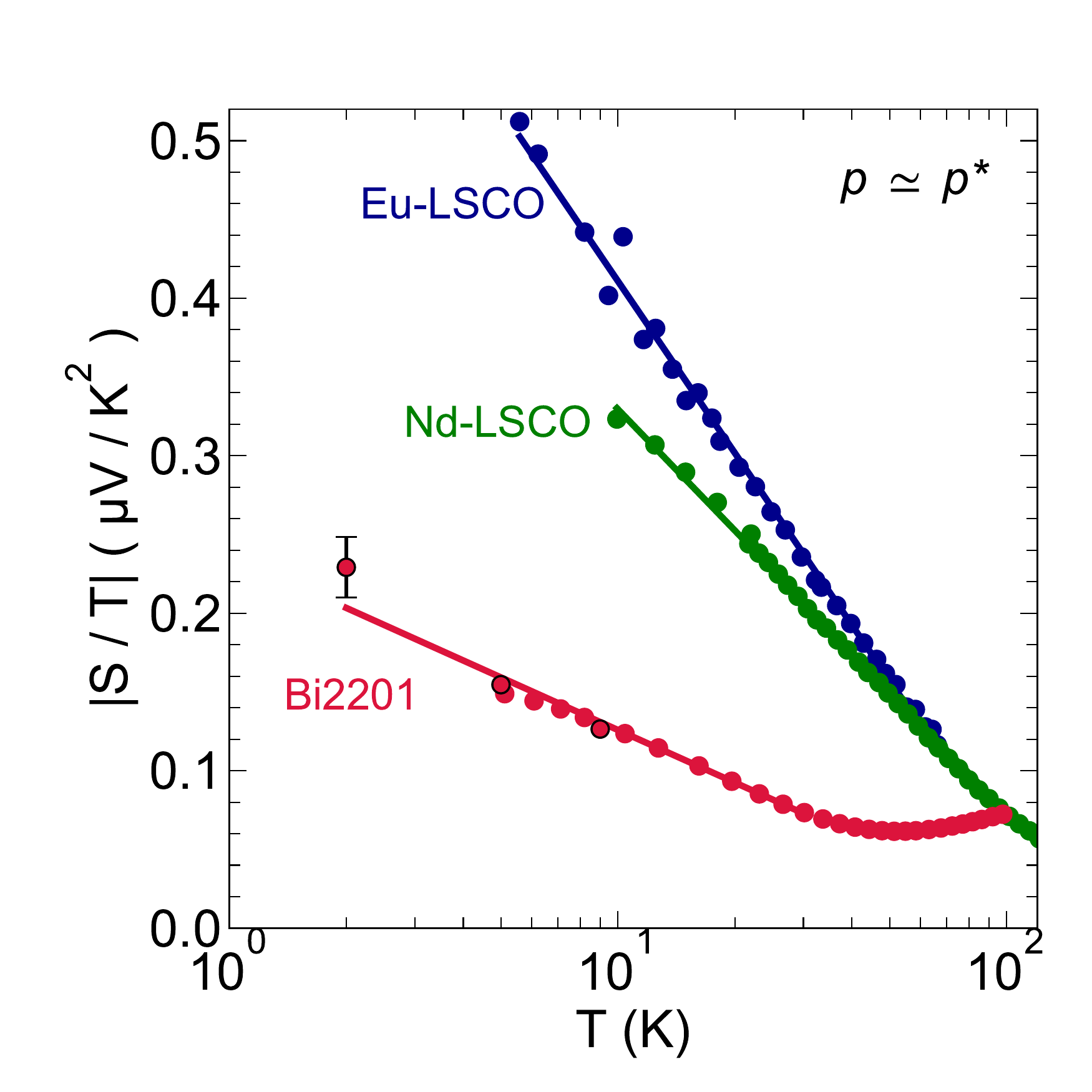}
\caption{
Temperature dependence of the normal-state Seebeck coefficient (absolute value) in three cuprates close to their respective
pseudogap critical points, plotted as $|S/T|$ vs $T$ on a semi-log plot:
Bi2201 with $p = 0.39$ (red, this work, OD10K) at $H = 15$~T (red dots) and $H = 20$~T (red dots with black contour), where \pstar~$= 0.40$;
Nd-LSCO with $p = 0.24$ at $H = 15$~T (green, ref.~\onlinecite{DaouPRB2009}), where \pstar~$= 0.23$; 
and 
Eu-LSCO with $p = 0.24$ at $H = 10$~T (blue, ref.~\onlinecite{laliberte2011fermi}), where \pstar~$= 0.23$.
Lines are a linear fit to the low-$T$ data.
We see that in all cases the behaviour at $T \to 0$ is $|S/T| \sim \log(1/T)$.
} 
\label{Fig.S-vs-logT_comparison}
\end{figure}

In Fig.~\ref{Fig.Seebeck}(a) and (b),
we see that $S$ remains negative down to the lowest temperature.
Strikingly, we observe a strong increase in the magnitude of $S/T$ upon cooling, 
with $|S/T|$ nearly doubling between 9~K and 2~K, 
which suggests a divergence as $T \to 0$. 
In Fig.~\ref{Fig.Seebeck}(c) and (d),
we see that $|S/T| \sim \log(1/T)$, from $T \simeq 40$~K down to at least 2~K.
Such a $\log(1/T)$ divergence of the Seebeck coefficient as $T \to 0$ has been observed in other cuprates.
In Fig.~\ref{Fig.S-vs-logT_comparison}, 
we compare our data in Bi2201 (OD10K) to published data in Nd-LSCO with $p=0.24$ (at $H=15$~T)~\cite{DaouPRB2009} 
and in Eu-LSCO with $p=0.24$ (at $H=10$~T)~\cite{laliberte2011fermi}.
We observe a $\log(1/T)$ behaviour in all three over a comparable temperature range.

In Eu-LSCO and Nd-LSCO, 
a $\log(1/T)$ divergence
is also observed in the electronic specific heat, $(C_{el}/T)$, at \pstar~\cite{michon2019thermodynamic}. 
This thermodynamic evidence of quantum criticality, combined with the $T$-linear resistivity,
suggests that 
the $\log(1/T)$ dependence of the Seebeck coefficient is also a signature of quantum criticality,
as found in some theoretical models~\cite{Paul2001}.
Note that in Nd-LSCO and Eu-LSCO the $\log(1/T)$ dependence of $S$ was observed at $p = 0.24$,
so just {\it above} \pstar~$=0.23$, 
whereas in our samples of Bi2201, we observe this $\log(1/T)$ dependence of $S$ at $p = 0.38$ and $p = 0.36$,
just {\it below} \pstar~$=0.40$ (Fig.~1).
Therefore, if this logarithmic dependence of the Seebeck coefficient is a signature of quantum criticality, it is an unconventional one, 
reminiscent of the specific heat of Nd-LSCO and Eu-LSCO,
in which the $\log(1/T)$ dependence of $C_{\rm el}/T$
is seen not only at $p = 0.24 >$~\pstar, but also at  $p = 0.22$ in Nd-LSCO and $p = 0.21$ in Eu-LSCO,
so at $p <$~\pstar~as well~\cite{michon2019thermodynamic}.
We conclude that the observation of a $\log(1/T)$ divergence in the Seebeck coefficient of Bi2201 makes a strong case for the universality of this signature, reminiscent of quantum criticality, among hole-doped cuprates at the pseudogap critical point.


\section{Thermal Hall conductivity: New signature of the pseudogap phase} 


A new signature of the pseudogap phase has recently been discovered in Nd-LSCO and Eu-LSCO: a large negative thermal Hall signal appears suddenly as the doping is reduced below \pstar, and it is entirely absent above \pstar~\cite{Grissonnanche2019}.
The low-temperature trend of this thermal Hall conductivity \Kxy~cannot be explained by the simple motion of charged carriers, since \Kxy~and $R_H$ have opposite signs as $T\rightarrow$0.
This large negative \Kxy~also persists down to $p = 0$, as confirmed in three cuprate Mott insulators -- La$_2$CuO$_4$, Nd$_2$CuO$_4$, and Sr$_2$CuO$_2$Cl$_2$~\cite{Boulanger2020}.
The phenomenon has since been attributed to phonons~\cite{Grissonnanche2020}, because an equally large signal was seen for a current along the $c$ direction, i.e. \Kzy~$\simeq$~\Kxy, and phonons are the only excitations in cuprates that travel as easily in directions parallel and perpendicular to the CuO$_2$ planes, as shown from the fact that \Kzz~$\simeq$~\Kxx~in La$_2$CuO$_4$ at low $T$~\cite{Hess2003}.
But since phonons are not sensitive to magnetic field on their own, they must be coupled to it through some mechanism (here the expression \textit{chiral phonons} stands for this sensitivity to a magnetic field).

\begin{figure}[t]
\centering
\includegraphics[width=0.93\columnwidth]{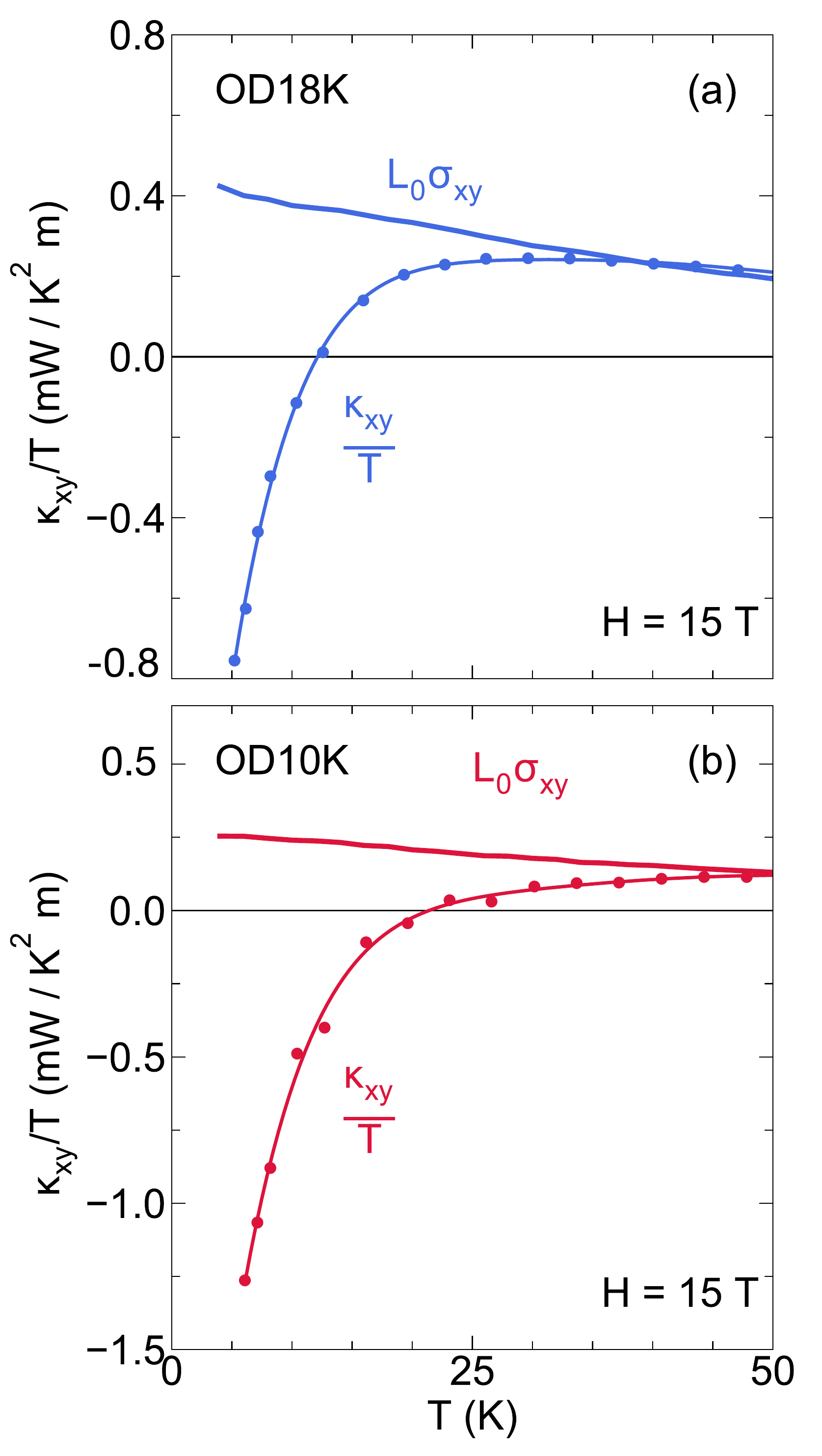}
\caption{
Thermal Hall conductivity $\kappa_{xy}$ of Bi2201 measured at $H = 15$~T, plotted as $\kappa_{xy}/T$ vs $T$, for our two samples:
(a) 
OD18K (blue dots); 
(b)
OD10K (red dots).
The thin lines are a guide to the eye through the data points.
In both panels, we also show the electrical Hall conductivity, 
$\sigma_{xy} = \rho_{xy} / (\rho_{xx}^2 + \rho_{xy}^2)$, 
plotted as $L_0 \sigma_{xy}$ vs $T$ (thick lines),
calculated using the data in Fig.~\ref{Fig.Hall}(c) ($\rho_{xy} \equiv$ \RH$(T) \times H$) and in Fig.~\ref{Fig.squid} ($\rho_{xx} \equiv \rho(T)$). 
} 
\label{Fig.Kxy}
\end{figure}

In Fig.~\ref{Fig.Kxy},
we show the thermal Hall conductivity of our two Bi2201 samples,
plotted as \Kxy$/T$ vs $T$.
A very similar qualitative behaviour is observed in both.
(Note that our data on OD18K was already reported in ref.~\onlinecite{Grissonnanche2019}, but not our data on OD10K.)
At $T = 50$~K, \Kxy~is positive and the thermal and electrical Hall conductivities are equal,
{\it i.e.} \Kxy$/T \simeq L_0 \sigma_{xy}$, showing that \Kxy~is entirely coming from the charge carriers
that are responsible for the electrical Hall effect and the resistivity.
Note that the thermal Hall effect at $T = 50$~K provides an independent measurement of how the Hall number changes between
OD10K and OD18K,
since \Kxy$/T \propto L_0/$\nH. We see that \Kxy~at $T = 50$~K increases by a factor 1.8, the same factor by which \RH$(0)$~increases
(Fig.~\ref{Fig.Hall}(c)), so that the drop in \nH~by a factor 1.8 obtained from electrical measurements~(Fig.~\ref{Fig.nH-vs-p}(a))
is confirmed by thermal measurements.

As temperature is reduced, \Kxy~changes sign and becomes increasingly negative as $T \to 0$ (Fig.~\ref{Fig.Kxy}).
By contrast, $L_0 \sigma_{xy}$ remains positive and just increases slowly and monotonically as $T \to 0$.
The negative component of \Kxy~is clearly not due to charge carriers, and phonons are the most likely carriers.
The fact that the magnitude of \Kxy~scales with the magnitude of phonon-dominated \Kxx~in the cuprate Mott insulators 
supports the view that phonons are responsible for the negative \Kxy~\cite{Boulanger2020}.
At $T \simeq 20$~K, the ratio $|$\Kxy$| /$\Kxx~is equal to 
0.30\% in La$_2$CuO$_4$, 
0.37\% in Nd$_2$CuO$_4$, 
and 
0.26\% in Sr$_2$CuO$_2$Cl$_2$~\cite{Boulanger2020}, 
and the ratio $|$\Kzy$| /$\Kzz~is equal to
0.48\% in Nd-LSCO at $p = 0.21$.
In our Bi2201 samples, 
we estimate the phonon term in the two conductivities as
\Kxx$^{\rm ph} = \kappa_{xx} - L_0 \sigma_{xx} T$
and 
\Kxy$^{\rm ph} = \kappa_{xy} - L_0 \sigma_{xy} T$,
and we obtain the ratio 
$| \kappa_{xy}^{\rm ph} /  \kappa_{xx}^{\rm ph}| \simeq 0.38$\% at $T = 5$~K and $H = 15$~T, for both samples.
This ratio is comparable to that found in the other cuprates~\cite{Boulanger2020}, pointing to a common mechanism.
The question is:
what property of the pseudogap phase makes phonons become chiral below \pstar~in hole-doped cuprates?

\section{Summary} 

We have measured five transport properties of the cuprate material Bi2201 in magnetic fields large enough to suppress superconductivity and reach the normal state in the $T = 0$ limit:
the electrical resistivity $\rho$, 
the electrical Hall coefficient \RH,
the thermal conductivity $\kappa$,
the Seebeck coefficient $S$,
and
the thermal Hall conductivity \Kxy.
Two dopings just below the pseudogap critical point \pstar~$= 0.4$ were investigated,
namely $p = 0.36$ and 0.39 (the absolute values of the hole doping are here estimated from ARPES measurements and do not impact the main conclusions that are drawn from this work, since it is the position of the samples relative to \pstar~that matters).
For the doping closest to \pstar,
we observe a $T$-linear resistivity down to 
\Tstar~$\simeq 25$~K,
with a slope per CuO$_2$ plane that is consistent with Planckian dissipation,
along with a logarithmically diverging Seebeck coefficient, $S/T$ $\propto$ log(1/$T$).
These two properties are typical signatures of quantum criticality.
We also observe a significant drop in the Hall number with decreasing doping, consistent with a rapid loss of carrier density below \pstar, concomitant with a drop in the density of states measured by the NMR Knight shift.
The Wiedemann-Franz law is satisfied, confirming that the normal state of the pseudogap phase is metallic. 
Finally, the thermal Hall conductivity shows a large negative contribution that we attribute to phonons.
The addition of Bi2201 to the list of cuprate materials in which these properties were already observed,
either in total (Nd-LSCO) or in part (Eu-LSCO, Bi2212, YBCO, Tl2201),
makes a strong case for the universality of the transport signatures of the pseudogap phase in hole-doped cuprates.\\


\section{Acknowledgments} 

L.T. acknowledges support from the Canadian Institute for Advanced Research (CIFAR) as a CIFAR Fellow
and funding from the Natural Sciences and Engineering Research Council of Canada (NSERC; PIN:123817), 
the Fonds de recherche du Qu\'ebec -- Nature et Technologies (FRQNT), 
the Canada Foundation for Innovation (CFI), 
and a Canada Research Chair. 
C.P. acknowledges support from the EUR grant NanoX n$^\circ$ANR-17-EURE-0009 and from the ANR grant NEPTUN n$^\circ$ANR-19-CE30-0019-01.
This research was undertaken thanks in part to funding from the Canada First Research Excellence Fund. 
Part of this work was funded by the Gordon and Betty Moore Foundation's EPiQS Initiative (Grant GBMF5306 to L.T.).
Part of this work was performed at the LNCMI, a member of the European Magnetic Field Laboratory.


\bibliography{Lizaire-Bi2201-17sept2021}

\end{document}